# Ideal antiferroelectricity with large digital electrostrain in PbZrO$_3$ epitaxial thin films

Yangyang Si[1,#], Ningbo Fan[2,#], Yongqi Dong[3,#], Zhen Ye[4,5,#], Shiqing Deng[6,7], Yijie Li[1], Chao Zhou[1], Qibin Zeng[4], Lu You[3], Yimei Zhu[7], Zhenlin Luo[2], Sujit Das[8], Laurent Bellaiche[9], Bin Xu[3,*], Huajun Liu[4,*], Zuhuang Chen[1,*]

1. State Key Laboratory of Advanced Welding and Joining of Materials and Structures, School of Materials Science and Engineering, Harbin Institute of Technology, Shenzhen, 518055, China
2. Jiangsu Key Laboratory of Thin Films, School of Physical Science and Technology, Soochow University, Suzhou 215006, China
3. National Synchrotron Radiation Laboratory, University of Science and Technology of China, Hefei, Anhui 230026, China
4. Institute of Materials Research and Engineering (IMRE), Agency for Science, Technology and Research (A*STAR), 2 Fusionopolis Way, Innovis #08-03, Singapore 138634, Republic of Singapore
5. Department of Mechanical Engineering, National University of Singapore, 9 Engineering Drive 1 Singapore 117575, Republic of Singapore
6. Beijing Advanced Innovation Center for Materials Genome Engineering, University of Science and Technology Beijing, Beijing, 100083, P.R. China
7. Condensed Matter Physics and Materials Science Department, Brookhaven National Laboratory, Upton, NY 11973, USA
8. Materials Research Centre, Indian Institute of Science, Bangalore 560012, India
9. Physics Department and Institute for Nanoscience and Engineering, University of Arkansas, Fayetteville, AR 72701, USA

[#]These authors contributed equally to this work.
**Corresponding authors:** zuhuang@hit.edu.cn; binxu19@suda.edu.cn; liu_huajun@imre.a-star.edu.sg

**Abstract**

Antiferroelectrics exhibit reversible antipolar-polar phase transitions under electric fields, yielding large electrostrain suitable for electromechanical devices. Nevertheless, in thin-film form, the antiferroelectric behavior is often obscured by competing ferroic orders, resulting in slanted hysteresis loops with undesired remnant polarization, subsequently posing challenges in obtaining ideal antiferroelectricity and understanding their intrinsic electrical behavior. Here, atomistic models for controllable antiferroelectric-ferroelectric phase transition pathways are unveiled along specific crystallographic directions. Guided by the anisotropic phase transition and orientation design, we achieved ideal antiferroelectricity with square double hysteresis loop, large saturated polarization (~60 $\mu C/cm^2$), near-zero remnant polarization, fast response time (~75 ns), and near-fatigue-free performance (~$10^{10}$ cycles) in $(111)_P$-oriented $PbZrO_3$ epitaxial thin films. Moreover, a bipolar and frequency-independent digital electrostrain (~0.83%) were demonstrated in this architype antiferroelectric system. In-situ X-ray diffraction studies further reveal that the large digital electrostrain results from intrinsic field-induced antiferroelectric-ferroelectric structural transition. This work demonstrates the anisotropic phase transition mechanism and ideal antiferroelectricity with large digital electrostrain in antiferroelectric thin films, offering a new avenue for applications of antiferroelectricity in nanoelectromechanical systems.

Materials exhibiting high electromechanical responses are of significant interest in modern nano/micro-electromechanical system [1-3]. Generally, there are two different types of electromechanical responses: analog type, where the induced displacement changes continuously with varying an applied field; and digital type, characterized by a critical electric field to turn on/off the electrostrain[4] (Fig. 1a). Analog displacements and electrostrains, commonly observed in piezoelectric and electrostrictive materials, offer continuous adjustment and have been extensively utilized in applications such as energy harvesting, displacement transducers, and ultra-precision positioners[5, 6]. The digital counterpart, in contrast, can provide discrete movement with a constant distance, *i.e.*, bistable on/off electrostrain states, which is well suited for applications in optical-grid manufacturing and swing motion devices[7]. However, achieving digital displacement behaviors is challenging in conventional electromechanical coupling materials because of the requirements of near-zero electromechanical response and significant displacement switching below and above the critical electric field, respectively.

Benefiting from the small energy difference between the ferroelectric (FE) and antiferroelectric (AFE) structures, AFEs are widely recognized as typical two-state systems where nonpolar and polar states as well as corresponding physical properties can be reversibly switched via application/removal of an external electric field[8, 9]. Consequently, it has emerged as an excellent candidate for efficient electric control of physical properties in modern electronic devices, and has been demonstrated towards various applications, *e.g.*, thermal switching[10], electrocaloric cooling[11], negative capacitance[12] and pulsed power capacitors[13]. Considering the large structural variations, the AFE-to-FE phase transition leads to substantial volume expansion under an applied electric field, resulting in large electromechanical responses (Fig. 1b)[7]. Since the field-induced electrostrain in the AFE or FE phase is generally negligibly small compared to the large electrostrain at the phase transition, therefore, for a sharp transition, a large digital-like electrostrain would be generated. This is distinctively different from the normal piezoelectric and electrostrictive effect. However, a comprehensive understanding of the AFE-to-FE phase transition is still lacking, and challenges to synthesize high-quality material limit the exploration of antiferroelectricity in both fundamental science and practical applications [14-16].

Meanwhile, taking archetypal AFE material $PbZrO_3$ (PZO) as an example, in its thin-film form, strain and/or surface effects would generally lead to the emergent FE order[17, 18]. As a result, antiferroelectricity in thin films is usually obscured with large remnant polarization and a slanted hysteresis loop[18]. Moreover, a new ferrielectric structure has been observed recently in $(100)_P$-oriented PZO films, indicating that there could be a transient state between the AFE and FE structures under an electric field[19, 20]. The existence of transient polar states during the AFE-FE phase transition would disrupt the digital signal. All told, a complete understanding of the antiferroelectricity is hardly realized, and direct evidence of ideal antiferroelectricity with sharp switching is still absent in AFE thin films[18-21]. The retention of ferroelectric component and the resulting remnant polarization strongly distorts the double hysteresis loop and hinders the digital electromechanical behavior in AFE thin films. The design and experimental proof of a sharp AFE-to-FE phase transition in thin films with large digital electrostrain remain to be explored.

In this work, we unveil the anisotropic antiferroelectric phase transition by combining first-principles calculations and experiments that demonstrate the presence of "ideal" antiferroelectricity with sharp phase transition and a large digital electrostrain in (111)$_P$-oriented PZO epitaxial thin films. The achieved double hysteresis loop exhibits a square shape with sharp switching, large saturated polarization (~60 μC/cm$^2$), near-zero remanent polarization, and fatigue-free behavior after $10^{10}$ switching cycles. Benefiting from the structural phase transition with a narrow voltage range, we achieve a large electrostrain (~0.83%). A field-induced FE structure is further resolved by in-situ synchrotron-based X-ray diffractions, providing the evidence of volume expansion with rhombohedral symmetry as long-term expected by communities. The current work makes a significant breakthrough for offering a holistic understanding of antiferroelectric behavior from structural to electrical phase transition, realizing sharp antiferroelectric switching in epitaxial thin films, along with digital electromechanical characterization and applications.

**Atomistic modelling of AFE-FE phase transition**

In its bulk form, PZO possesses an orthorhombically distorted perovskite structure with space group *Pbam* and lattice constant of $a_o$ = 5.884 Å, $b_o$ = 11.768 Å, and $c_o$ = 8.220 Å at room temperature[22]. The orthorhombic unit cell of PZO can alternatively be described as a simplified pseudotetragonal perovskite unit cell with constants $a_p = a_o/\sqrt{2}$ = 4.16 Å, $b_p = b_o/2\sqrt{2}$ = 4.16 Å and $c_p = c_o/2$ = 4.11 Å[22] (The subscript "O" denotes orthorhombic indices while the subscript "*P*" denotes pseudo-tetragonal indices, Supplementary Fig. 1). The centrosymmetric orthorhombic structure is associated with an $a^-a^-c^0$ oxygen octahedral rotation pattern in Glazer's notation[23], together with Pb$^{2+}$ cations adopting the up-up-down-down (*i.e.*, ↑↑↓↓) antipolar dipole arrangements along [110]$_P$. In comparison, the field-induced FE structure is generally predicted to be the rhombohedral *R*3*c* structure with polarization direction along [111]$_P$[24], despite the fact that its exact lattice structure has not been experimentally validated yet. Fig. 1c shows the AFE structure alongside the FE structure of PZO, both being relaxed from DFT calculations in Supplementary Table 1. Compared with the *Pbam* AFE structure, wherein the antipolar dipoles are aligned along [110]$_P$, the dipoles in the FE structure switch to the [111]$_P$ direction, and additional antiphase rotation occurs along the [001]$_P$ axis, which give rise to $a^-a^-a^-$ octahedral rotation. Therefore, both dipole reorientation and antiphase octahedral rotation play important roles in the AFE-to-FE phase transitions, and strong anisotropy is expected when the antipolar dipoles are subjected to external electric fields with different directions. Considering the orientation geometry, DFT calculations are performed under applied electric fields of four different directions, *i.e.*, [100]$_P$, [110]$_P$, [1$\bar{1}$0]$_P$ and [111]$_P$ directions. The enthalpy (defined as $H = E_{\text{DFT}} - P \cdot E$, where $E_{\text{DFT}}$ is the DFT energy, $P$ is the ferroelectric polarization, and $E$ is the applied electric field) gradually decreases with increasing electric field before the AFE-to-FE phase transition in Fig. 1d. Sharp enthalpy change occurs at the critical field where the AFE structure transforms into the FE structure. The enthalpy of the FE structure is lower than that of the AFE under high-field conditions, which can be understood as the driving force of the AFE-to-FE phase transition.

The variation in polarization and volume of the AFE structure under an electric field are further calculated (Supplementary Fig. 1). Consequently, PZO under a [111]$_P$-oriented electric field exhibits the largest polarization change, along with the smallest critical transition field, highlighting the advantages of orientation control in manipulating antiferroelectricity. Meanwhile, the final FE structure obtained under applied electric fields of different directions show different symmetries (Supplementary Fig. 2). Only the PZO under the [111]$_P$ field exhibits the rhombohedral structure with *R*3*c* symmetry which has the largest volume expansion, while the other three are monoclinic structures with *Cc* symmetry and smaller volume expansion than that of the [111]$_P$ case shown in Fig. 1e. This difference can be attributed to the non-parallel alignment between the field direction and polarization vector in the field-driven FE phase, which can distort the polarization orientation and destroy the three-fold rotation symmetric around the [111]$_P$ axis in the *R*3*c* phase, leading to the formation of the lower-symmetry *Cc* phase for fields along the [100]$_P$, [110]$_P$, and [1$\bar{1}$0]$_P$ directions.

We further performed nudged elastic band (NEB) calculation to identify the structural evolution and possible transient states during the AFE-FE phase transition with different field directions. The energetic pathways of PZO related to the reaction coordinate is presented in Fig. 1f. Two local minima appear when the electric field is applied along the [1$\bar{1}$0]$_P$ and [100]$_P$ directions, indicating the existence of two additional metastable transient states IM* and IM** in Fig. 1g. In contrast, no local minimum appears when the applied field is along the [111]$_P$ and [110]$_P$ directions. From the perspective of atomic structure, antiphase rotation and antipolar dipole switching occurs simultaneously in PZO under the application of electric field along the [111]$_P$ direction (Supplementary Fig. 3). However, in the atomic structure of PZO under field along the [1$\bar{1}$0]$_P$ and [100]$_P$ directions, the antipolar dipoles can partially switch first from ↑↑↓↓ in AFE to a transient ferrielectric state (IM*) with ↑↑↓↑ polar arrangement in the (001) plane, accompanying with the presence of polarization component and antiphase octahedral rotation along the [001]$_P$ direction (Supplementary Fig. 4). Subsequently, they fully switched to another transient state (IM**) with ↑↑↑↑ polar arrangement, which is close to the [111]$_P$ direction. The final FE structure can be further achieved from the polar transient state IM** through octahedral rotation-assisted polar dipole reorientation to the [1$\bar{1}$1]$_P$ direction. The complex structural evolution with multiple switching differs significantly from the one-step AFE-to-FE phase transition in PZO under application of fields along the [111]$_P$ direction, and can result in a deviation from the expected one-step AFE behavior. All told, our DFT results demonstrate that sharp transition can be derived from PZO with appropriate applied field direction (*i.e.*, [111]$_P$ direction), resulting in the sharpest switching event, largest saturated polarization and volume expansion.

**Antiferroelectric structure of (111)$_P$ -oriented PbZrO$_3$ thin films**

To examine the theoretical predication, antiferroelectric PZO thin films with thickness of ~150 nm is epitaxially grown on SrRuO$_3$ (SRO)-buffered (111)$_P$ -oriented SrTiO$_3$ (STO) single crystal substrates via pulsed laser deposition. X-ray diffraction analysis is performed to analyze crystal structure in PZO films, revealing the high-quality epitaxy of the film without any secondary phases in the *θ/2θ* scan in Fig. 2a. Apart from the main Bragg diffraction peaks from PZO, SRO, and STO, a superlattice peak corresponding to the (3/2, 3/2, 3/2)$_P$ reflection of PZO also appears. This half-index peak is forbidden in antiphase octahedral rotation but emerges with accompanying of antipolar dipole arrangement[25]. For (111)$_P$ -oriented PZO-based AFE films, our previous studies revealed that the AFE axis points along the [110]$_P$ direction and thus inclines by 35.26° from the out-of-plane [111]$_P$ direction[26]. Therefore, in-plane antipolar modulation would orient along [1$\bar{1}$0]$_P$ and subsequently result in 60° domains (inset of Fig. 2a). Therefore, X-ray RSMs are further conducted to study the domain structure in Fig. 2b. Clear 3-fold splitting of the PZO 111$_P$ Bragg peak was observed in the *HK* mapping, confirming the existence of 60° structural domains. The triple splitting can be attributed to the pseudo-tetragonal structure of PZO as illustrated in Supplementary Fig. 6 and Supplementary Fig. 7. In addition, the presence of 60° domains further leads to the hexagonal satellite 1/4 superstructure diffraction peaks originated from the antipolar dipole modulation.

High-angle annular dark-field (HAADF) scanning transmission electron microscopy (STEM) imaging is employed at the $[11\bar{2}]_P$ zone axis to directly observe antipolar dipole arrangement in real space. The cross-sectional image of PZO displays high crystalline quality without any visible defect in Fig. 2c. The atomic model fits well with the distribution of atoms from the HAADF-STEM results in Fig. 2d, where $Zr^{4+}$ are neatly aligned while $Pb^{2+}$ display antiparallel displacement relative to the centrosymmetric $Zr^{4+}$ positions. The corresponding fast Fourier transform reveals the presence of 1/4 superstructure spots along $[1\bar{1}0]_P$, which agrees well with the in-plane ↑↑↓↓ antipolar dipole arrangement in Fig. 2e. Albeit with some disturbance originating from the partially overlapping between $Pb^{2+}$ and $Zr^{4+}$, the antipolar dipoles and their arrangement can be effectively observed across $[11\bar{2}]_P$ zone axis (Supplementary Fig. 8), and the antipolar nature with ↑↑↓↓ of $Pb^{2+}$ displacement is clearly visible. All told, the atomic-scale real-space imaging studies clearly reveal the antiferroelectric structure in the $(111)_P$-oriented PZO thin films, being consistent with aforementioned X-ray diffraction results.

**Antiferroelectric switching under electric field**

For ideal antiferroelectricity, the AFE-FE phase transition under electric field should be first order in a narrow electric field range, that is, a sharp phase transition with large saturated polarization and zero remanent polarization is preferred. In the $(111)_P$-oriented PZO thin films, a square double hysteresis loop with large saturated polarization ($P_s$ ~ 60 μC/cm$^2$) and four sharp switching current peaks is obtained in Fig. 3a. The double hysteresis loop shows nearly zero remanent polarization and a steep slope at the critical electric field, which is almost perpendicular to the electric field axis. In comparison, a slanted quintuple hysteresis loop with lower saturation polarization ($P_s$ ~ 40 μC/cm$^2$) and larger remanent polarization ($P_r$ ~ 2.9 μC/cm$^2$) are observed in the $(100)_P$-oriented PZO thin films (inset of Fig. 3a). Meanwhile, the switching current exhibit ten switching peaks, indicating gradual multistep phase transitions between the AFE and FE structure, which is well consistent with aforementioned NEB results. The sharp switching in the $(111)_P$-oriented PZO thin film not only prevents the remanent polarization resulting from the competing ferroelectric order but also achieves the highest saturation polarization up to ~ 60 μC/cm$^2$, in good agreement with first-principal calculations.

Fatigue measurement in the $(111)_P$-oriented PZO film shows that the saturated polarization remains the same even after $10^{10}$ switching cycles in Fig. 3b. The observed fatigue-free behavior further implies high film quality with low defect concentration, as the presence of defects would pin dipole switching, consequently severely reduces the polarization value. The superior film quality can be further supported by the ultralow leakage current density (Supplementary Fig. 9). Apart from the appealing fatigue performance, the phase transition kinetics is further evaluated from pulse measurement in Fig. 3c. During a rapid increase in electric field, the polarization can quickly reach to the maximum within 75 ns. The results with different electrode sizes shows a monotonically scaling behavior between the transition speed and electrode area. Therefore, by decreasing the electrode area, the transition time would be further minimized to the sub-ns range, being comparable to typical ferroelectrics[27]. To well evaluate the superior antiferroelectricity, an antiferroelectric superior factor $\eta = (P_{max} - P_r) / P_r$ can be defined by combining $P_{max}$ and $P_r$, where the higher $P_{max}$ and lower $P_r$ would result in the higher $\eta$. Overall, the $(111)_P$ -oriented PZO thin films display nearly zero remanent polarization and the largest superior factor, demonstrating "ideal" antiferroelectricity as compared with other AFE thin films shown in Fig. 3d[3, 21, 28-44].

Based on the sharp transition process, digital displacement and electrostrain are expected in the $(111)_P$ -oriented PZO films for electromechanical application. To characterize the macroscopic electromechanical strain under an electric field, a laser scanning vibrometer (LSV) is used[45]. The displacement is relatively small (< 100 pm) below the transition field when the structure remains AFE in Fig. 4a. However, as the electric field increases, a uniform displacement jump as large as ~ 1100 pm is observed above the transition point (6V). The displacement profiles under different bias voltages are presented in Fig. 4b. Within a small voltage range, the displacement profiles show an abrupt change and saturate quickly after the transition point, which can be seen as digital electromechanical switching with a threshold field. This type of digital electromechanical switching is further reflected in the electrostrain versus electric field in Fig. 4c, which shows a steep slope at a narrow voltage gap and achieves large electrostrain up to ~0.83%. Moreover, due to the intrinsic field-induced AFE-FE transition, this digital electromechanical behavior shows bipolar feature in Fig. 4d and frequency independence from 500 Hz to 10000 Hz in Fig. 4e. This frequency-independence behavior arises from the first-order, field-driven structural phase transition. When the external electric field reaches the critical transition point, the phase transition occurs on a nanosecond timescale (Fig. 3c). Therefore, the digital electromechanical behavior exhibits broad frequency independence, provided that the measurement frequence does not exceed the switching dynamics limit.

**In-situ structural analysis of antiferroelectric and field-induced ferroelectric phase**

As the digital electrostrain response relies on the field-induced lattice change, it is essential to have a detailed comparison between the AFE and FE phases from perspectives of structure landscape. To obtain the evidence of structural symmetry change and accurate structural parameters, in-situ synchrotron-based 3D-reciprocal space mapping (RSM) is performed around PZO (222)$_P$ diffraction condition from both AFE and field-induced FE phases (Fig. 5a-b). Under zero field, the AFE structure displays a triple-split of the 222$_P$-diffraction spot due to its pseudo-tetragonal nature. Under an electric field higher than the critical point, a field-induced FE phase in PZO appears *i.e.*, the FE structure displays one symmetric 222$_P$-spot without splitting in Fig.5b. It's noteworthy that the $Q_z$ of the 222$_P$-spot from the FE structure is lower than that from AFE, which indicates the out-of-plane expansion during the AFE-to-FE phase transition. By computing the (222)$_P$ lattice spacing with $Q_z$, an out-of-plane lattice expansion can be obtained as ~ 0.77%, which is close to the electrostrain of ~ 0.83% measured by LSV, indicating the field-driven AFE-to-FE structural phase transition is the major contributor to the observed large digital displacement.

To further reveal the structure and symmetry differences between the AFE and FE phases, RSM studies, including the symmetric (222)$_P$ and other surrounding asymmetric main spots, are carried out. The splitting of main spots is clearly observed in the as-grown film, implying the existence of multidomain structures originating from the pseudo-tetragonal AFE structure. The three sets of main spots constitute the diffraction patterns in Fig. 5c and are differentiated by colors in Fig. 5d. The disappearance of main spots splitting under the applied field indicates the presence of field-induced highly symmetric single-domain FE structure in Fig. 5e and 5f. By reconstructing the lattice in real space from RSM, a rhombohedral structure can be concluded within the experimental accuracy in Supplementary Table 2, which support the long-standing predicted ferroelectric structure with rhombohedral symmetry in PZO.

In previous studies, the coexistence of heterophases at ground state and the presence of intermediate ferrielectric structure during AFE-to-FE phase transitions in (100)$_P$-oriented PZO thin films has been reported. These observations have led to ambiguous understandings on the antiferroelectricity and sparked ongoing debates about the intrinsic structure in PZO[21, 46]. In this study, by controlling the orientation of high-quality epitaxial PZO films, we were able to achieve sharp antiferroelectric switching in (111)$_P$-oriented films without any metastable ferrielectric/ferroelectric structure. The field-induced rhombohedral ferroelectric structure is observed for the first time by in-situ reciprocal space mapping and further supported by the DFT calculations. Meanwhile, NEB calculations suggest that the antipolar dipole switching are strongly anisotropic and correlated with the antiphase rotation along *c* axis, which alter the octahedral rotation from $a^-a^-c^0$ to $a^-a^-a^-$ in Glazer's notation and assisted the AFE-to-FE phase transition. These results provide a systematic analysis on the electrical and structural details in (111)$_P$-oriented PZO films, reveal the anisotropy effect on AFE thin films and emphasize the importance of electric field direction in determining the AFE behaviors.

**Conclusions**

In summary, we have demonstrated ideal antiferroelectricity with sharp phase transition in (111)$_P$-oriented PZO films. The observed large digital electrostrain, fast response time, and excellent fatigue performance pave the way towards digital control and electromechanical applications. The sharp antiferroelectric switching can be attributed to the anisotropic response of the AFE structure under applied electric field, where the antipolar dipole switching and antiphase octahedral rotation can concurrently develop in (111)$_P$-oriented PZO films. Our studies not only provide invaluable insight into the intrinsic behavior of PZO films, but also pave the way for the design of high-strain digital actuator applications using antiferroelectric thin films.

**Methods**

**Sample preparation.** High quality PbZrO$_3$/SrRuO$_3$ (PZO/SRO) heterostructures were synthesized on commercial (111)$_P$ -oriented SrTiO$_3$ (STO) single crystal substrates via pulsed laser deposition (PLD) (Arrayed Materials RP-B). The growth of SRO was accomplished at 680 ºC with a dynamic oxygen pressure of 100 mTorr. The corresponding laser condition is 1.0 J/cm$^2$ and 5 Hz for laser fluence and frequency, respectively. The growth of PZO was carried out at a lower temperature of 590 ºC with a dynamic oxygen pressure of 80 mTorr. The corresponding laser condition is 1.6 J/cm$^2$ and 5 Hz for laser fluence and frequency, respectively. After deposition, the samples were cooled down to room temperature at a rate of 5 °C min$^{-1}$ in an oxygen pressure of 700 Torr. After the growth, a standard lithography process is used to pattern circular photoresist pads on the heterostructure. After photolithography, the circular top Pt electrodes with areas ranging from 200 to 3200 μm$^2$ was deposited by magnetron sputtering (Arrayed Materials RS-M). The photoresist was then lifted off in acetone.

**X-ray diffraction characterization.** X-ray *θ-2θ* scans were obtained by high-resolution X-ray diffractometer (Rigaku Smartlab 9 KW). The in-situ 3D-RSM experiments under electric field were performed at Beamline BL02U2 at Shanghai Synchrotron Radiation Facility (SSRF, λ= 0.0680 nm), China. Pt electrodes with thickness of 100 nm and diameter of 500 μm were sputtered on the surfaces for in-situ XRD. The DC electric field was applied between bottom electrode SrRuO$_3$ and top electrode Pt using a high-voltage power supply. A series of diffraction patterns were collected by EigerX 500k pixel area detector with rotating the sample around θ axis with Δθ =0.02 steps from around 14.78° to 20.78°. The exposure time was 1 s per frame. The 3D reciprocal space maps were reconstructed with self-developed Python scripts.

**Aberration-corrected STEM characterization.** The scanning transmission electron microscopy (STEM) sample was prepared using focused ion beam milling. STEM investigations were conducted by using an aberration-corrected FEI Titan Themis G2 microscope operated at 300 kV, which was equipped with double aberration correctors and an X-FEG gun. The convergent semi-angle and collection angle were 25 mrad and 48 to 200 mrad for HAADF-STEM imaging, respectively. Quantitative treatments of HAADF-STEM images were performed by using a MATLAB program made in-house. The coordinate system was set as horizontal direction and vertical directions were X-axis and Y-axis, respectively, with the left top corner pixel being the coordinate origin.

**Electrical properties measurements.** The electrical measurements were performed with a TF3000 analyzer (aixACCT) on a Semishare high-precision probe station (Semishare E4). The P-E loop was characterized under a 1 kHz AC electric field. The fatigue performance was tested under a 1 kHz AC electric field after several electric cycles under a 100 kHz AC electric field at the same electrode. The cycle electric field was set to be 400 kV/cm, which is slightly higher than the critical electric field for antiferroelectric-ferroelectric phase transition under a 100 kHz AC electric field. After $10^{10}$ cycles of electric field training, the double hysteresis loop shows negligible degradation. The pulse test was conducted with Keithley 4200A-SCS Parameter Analyzer. A series of rectangular waves with electric field reaches up to 400 kV/cm was applied to induce AFE-to-FE phase transition. The corresponding risetime and pulse width for pulse measurement was set to be 20 ns and 5 us, respectively to distinguish nanosecond scale polarization switching dynamics.

**Electromechanical response Measurements.** Considering the potential impact of substrate bending during strain measurement, we use laser scan vibrometer to measure both the electrode and surrounding area without electrode, accounting for substrate bending effects. By subtracting the displacement of substrate ($\delta_{sub}$) from the displacement of the film under electrode ($\delta_{film}$), the effective displacement of the film can be determined by $\Delta\delta = \delta_{film} - \delta_{sub}$, thereby eliminating the contribution from substrate bending. This method effectively eliminates the influence of substrate bending [45, 47]. The electromechanical properties are measured by a laser scanning vibrometer (Polytec) system. Circular gold/palladium (with thicknesses of 50 nm and 30 nm) top electrodes of 200 μm in diameter were deposited on the PZO films by e-beam evaporator system, forming an Au/Pd-PZO-SRO sandwich device. A complex bias voltage was applied on the device through tungsten probes. The defined grid scanning area covered both the top electrode and surrounding areas without electrode. The effective piezoelectric coefficient is given by the slope in the field versus strain curve.

**DFT calculations.** First-principles density functional theory (DFT) calculations were performed using the Vienna ab initio simulation package (VASP)[48, 49]. The projector augmented wave (PAW) method was used for describing electron-ion interactions[50], and the generalized gradient approximation (GGA) parametrized by Perdew-Burke-Ernzerhof for solids (PBEsol) was used for the exchange-correlation functionals[51]. The orbitals of $5d^{10}6s^26p^2$, $4s^24p^64d^25s^2$, and $2s^22p^4$ were explicitly treated as valence electrons for Pb, Zr, and O, respectively. A cutoff energy of 500 eV for the plane-wave basis set and Γ-centered $8 \times 4 \times 6$ $k$-point mesh were used for the AFE (*Pbam* symmetry)[46]. We relaxed all atoms until the Hellmann-Feynman forces were less than 0.001 eV/Å. The convergence threshold of the self-consistent field iteration was set to $10^{-7}$ eV. The electrical polarizations were calculated based on the DFT calculated Born effective charges and atomic displacements with respect to the high-symmetry phase. And the structural response to the applied electric field was determined by considering the additional electrostatic force due to the Born charges, according to the scheme of Ref.[52]. Note that the Born charges were allowed to vary for different field strengths. Under the field, the equilibrium structural stability is determined by the electric enthalpy functional given by: $H = E_{\text{DFT}} - \boldsymbol{P} \cdot \boldsymbol{E}$, where $E_{\text{DFT}}$ is the DFT energy, $\boldsymbol{P}$ is the ferroelectric polarization, and $\boldsymbol{E}$ is the applied electric field. The energetic pathway and the associated barrier were computed by the climbing image nudged elastic band (CI-NEB) method[53]. Linearly interpolation between AFE and FE structures is adopted as the initial path, which is then optimized by CI-NEB until the Hellmann-Feynman forces were less than 0.02 eV/Å.

**Data availability**

All data supporting the findings of this study are available within the Article and Supplementary Information, and are available from the corresponding author on reasonable request.

**Code Availability**

The code used to calculate the results shown in this work is available from the corresponding authors upon reasonable request.

**Acknowledgements**

This work was supported by National Natural Science Foundation of China (Grant Nos. 52372105 and 92477129) and Shenzhen Science and Technology Program (Grant No. KQTD20200820113045083). Z.H.C. has been supported by State Key Laboratory of Precision Welding & Joining of Materials and Structures (Grant No. 24-Z-13) and "the Fundamental Research Funds for the Central Universities" (Grant No. 2024FRFK03012). Y.Q.D. and Z.L.L.

were supported by the National Natural Science Foundation of China (Grant No. 12175235, 12004407) and the National Key Research and Development Program of China (Grant No. 2022YFA1603902). N.B.F. and B.X. acknowledge financial support from National Natural Science Foundation of China (Grant No. 12074277), Projects of International Cooperation and Exchanges NSFC (Grant No. 12311530693), and the support from Priority Academic Program Development (PAPD) of Jiangsu Higher Education Institutions. H.J.L. acknowledges the support from National Research Foundation Competitive Research Program (NRF-CRP28-2022-0002), RIE2025 MTC Individual Research Grant (M22K2c0084), Career Development Fund (C210812020) and Central Research Fund, A*STAR, Singapore. L.B. thanks the Vannevar Bush Faculty Fellowship (VBFF) Grant No. N00014-20-1-2834 from the Department of Defense. The authors acknowledge the support from the BL02U2 of Shanghai Synchrotron Radiation Facility (SSRF).

**Author Contributions**

Z.H.C. and Y.Y.S. conceived of and designed the research. Z.H.C. supervise the research. Y.Y.S. grow the films, measured X-ray diffraction and electric performance. Y.Q.D, Z.L.L., Y.Y.S., and Y.J.L. carried out the In-situ 3D RSM measurement. N.B.F., K.P., L.B., and B.X. performed first-principles calculations. Z.Y., Q.B.Z., and H.J.L. performed laser scanning vibrometer measurement. S.Q.D. carried out STEM studies. Y.Y.S., Y.Q.D., N.B.F., Z.Y., B.X., H.J.L., and Z.H.C. analyzed the data. Y.Y.S., B.X., H.J.L. and Z.H.C. wrote the manuscript. All authors discussed and commented on the manuscript.

**Competing interests**

The authors declare no competing interests.

**Additional information**

The Supplementary Information is available free of charge on the website.

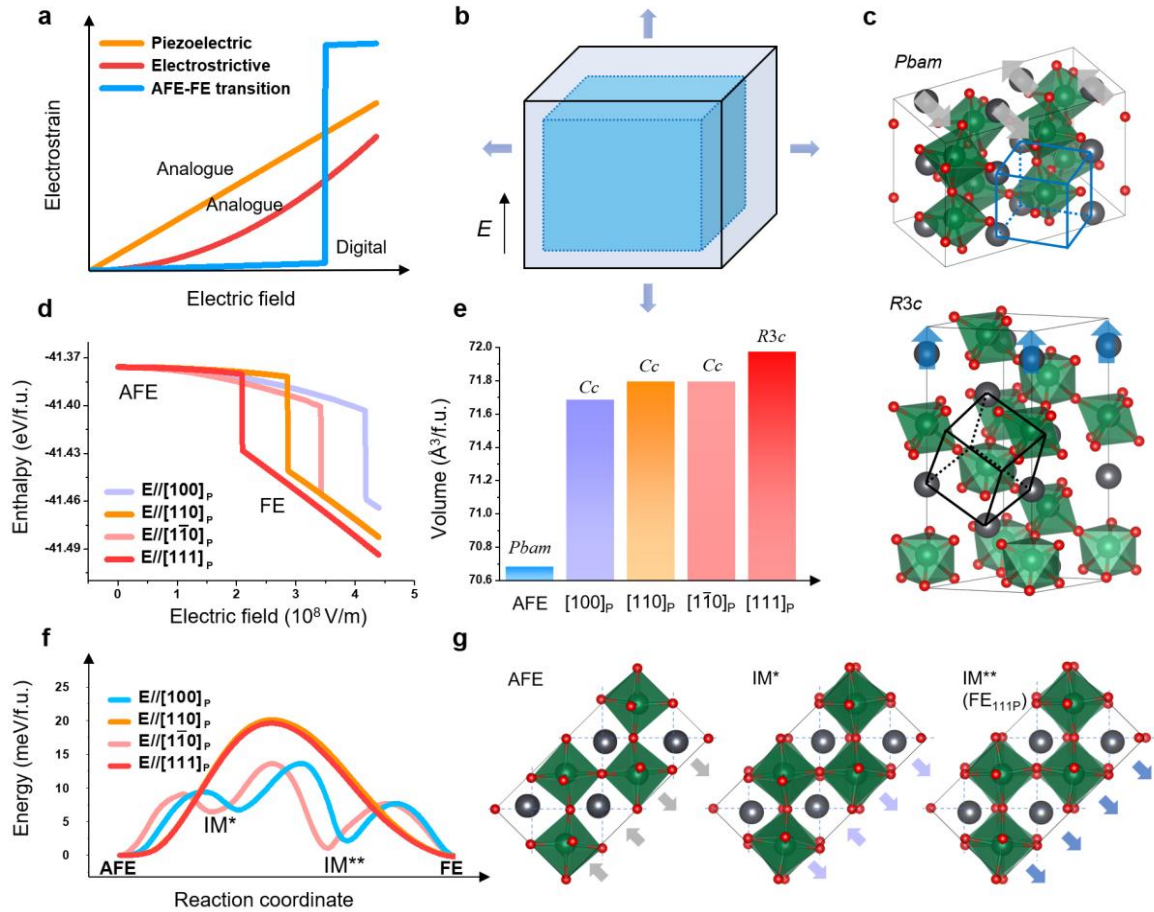

**Fig. 1 Atomistic modeling of AFE-FE phase transition in PbZrO₃. a,** Schematic comparison of electrostrain versus electric field among normal piezoelectric, electrostrictive and AFE-FE phase transition. **b,** Schematic view of volume expansion during AFE-FE phase transition. **c,** Structure of the AFE ground state and field-induced FE phase from DFT calculations, the atoms in grey, green and red denote the $Pb^{2+}$, $Zr^{4+}$ and $O^{2-}$, respectively. **d,** Anisotropic enthalpy and **e,** volume expansion during AFE-FE phase transition. **f,** Energetic path for AFE-FE phase transition with various electric field directions. **g,** AFE, IM* and IM** (FE₁₁₁ₚ) structures derived from NEB calculations.

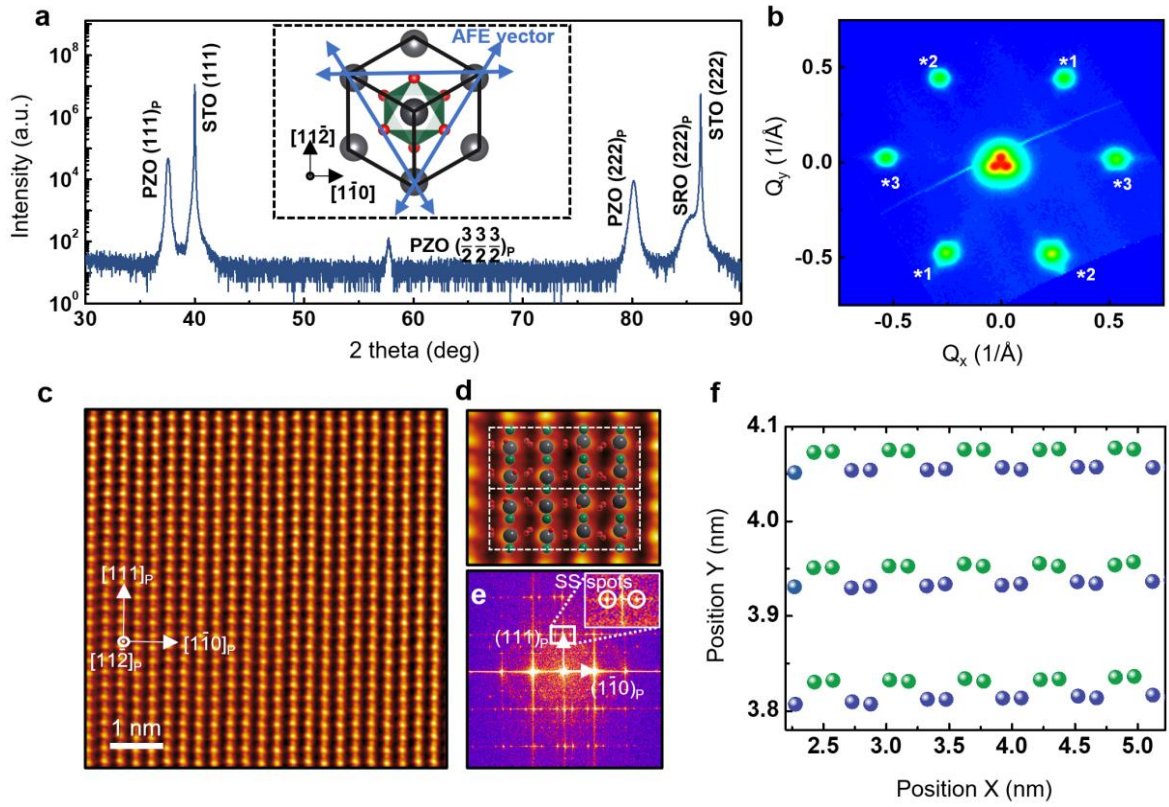

**Fig. 2 Antiferroelectric structure characterization in (111)$_P$-oriented PbZrO$_3$ thin films. a,** $\theta/2\theta$ line scan of 150 nm PbZrO$_3$/20 nm SrRuO$_3$/SrTiO$_3$ (111) heterostructure. Figure inset denotes the possible antiferroelectric vector orientations. **b,** Synchrotron-based in-plane reciprocal space mapping around (111)$_P$-diffraction condition of PbZrO$_3$ with $Q_x$//[1$\bar{1}$0]$_P$ and $Q_y$//[11$\bar{2}$]$_P$. The main Bragg spots of PbZrO$_3$ (111)$_P$ exhibit triple splitting, accompanied with hexagonal satellite spots. **c,** Cross-sectional HAADF-STEM image of PbZrO$_3$ layer. **d,** local AFE structure fitted by conventional atomistic model of *Pbam* PbZrO$_3$, the atoms in grey, green and red denote the Pb$^{2+}$, Zr$^{4+}$ and O$^{2-}$, respectively. **e,** Fast Fourier transform of PbZrO$_3$ layer in **c. f,** Selected Pb$^{2+}$ position mapping from dotted square in **c** reveals the antipolar period in PbZrO$_3$ layer.

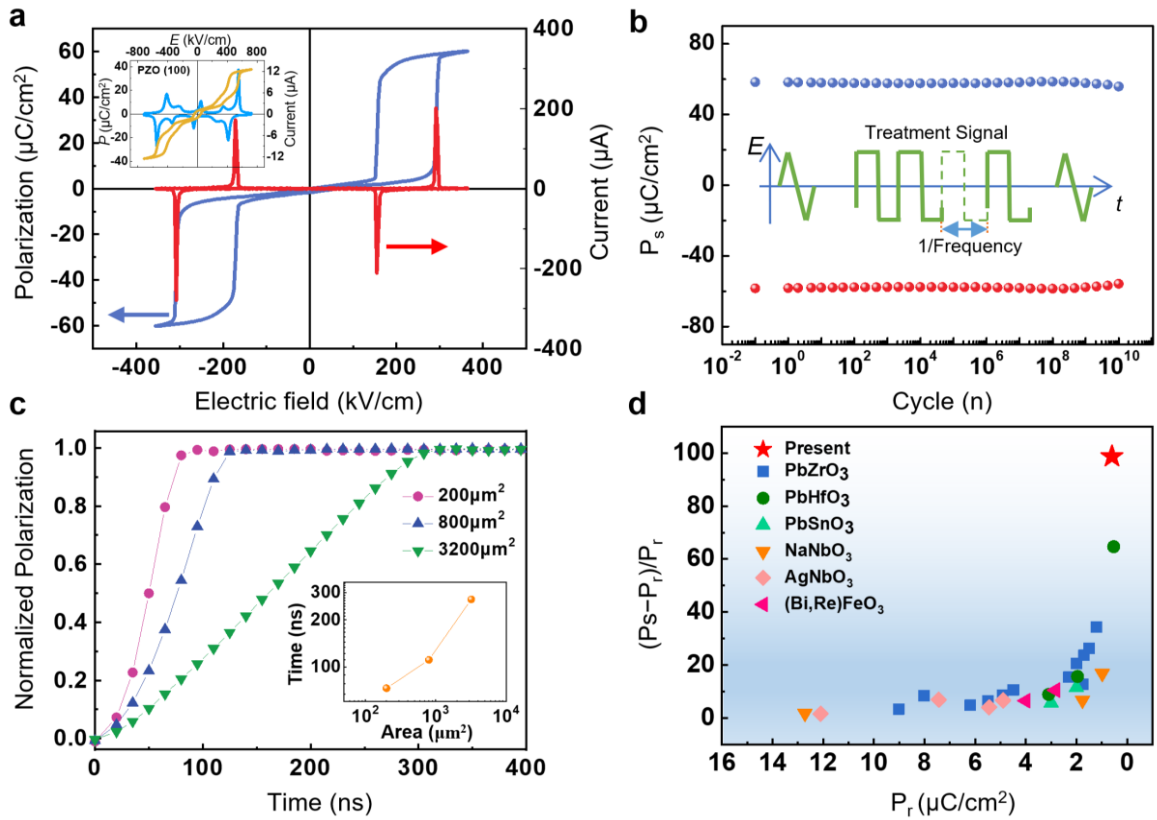

**Fig. 3 Electrical characterization of AFE-FE phase transition. a,** Double hysteresis loop and switching current of (111)$_P$-oriented PbZrO$_3$ films, figure inset provide the slanted double hysteresis loop and corresponding multistep switching current peak in (100)$_P$-oriented PbZrO$_3$ films. **b, c,** Fatigue performance **b** and switching time **c** for AFE-FE phase transition in (111)$_P$-oriented PbZrO$_3$ films. **d,** Comparison of superior factors and remanent polarization of our films with representative reported perovskite AFE thin films, including PbZrO$_3$ thin films[3, 21, 28-36], PbHfO$_3$ thin films[37, 38], PbSnO$_3$ thin films[39], (Bi,Re)FeO$_3$ thin films[40], AgNbO$_3$ thin films[41,42], and NaNbO$_3$ thin films[43,44].

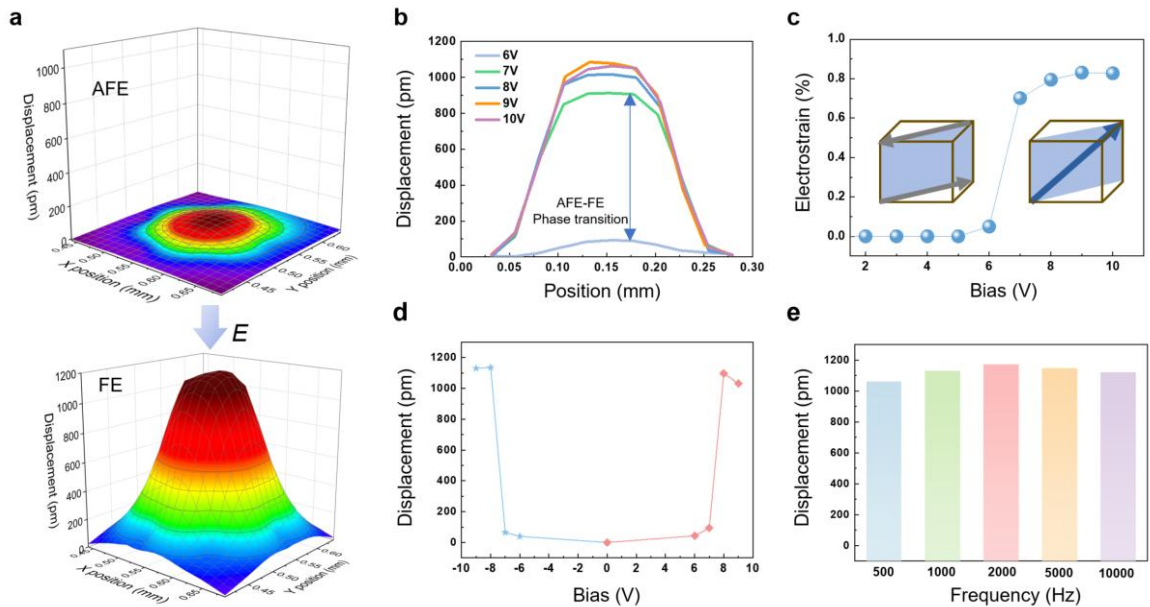

**Fig. 4 Electromechanical response during AFE-FE phase transition. a,** 3D drawing of surface displacements under applied electric field for AFE (upper panel) and FE (lower panel). **b,** Surface displacement profiles near the critical electrical field that induces the AFE-FE phase transition. **c,** The electrostrain evolution with increasing electric fields, figures inset represent the antipolar dipole geometry (left panel) and polar dipole geometry (right panel). **d,** Digital displacements under positive and negative bias. **e,** Frequency dependence of displacements during the AFE-FE phase transition.

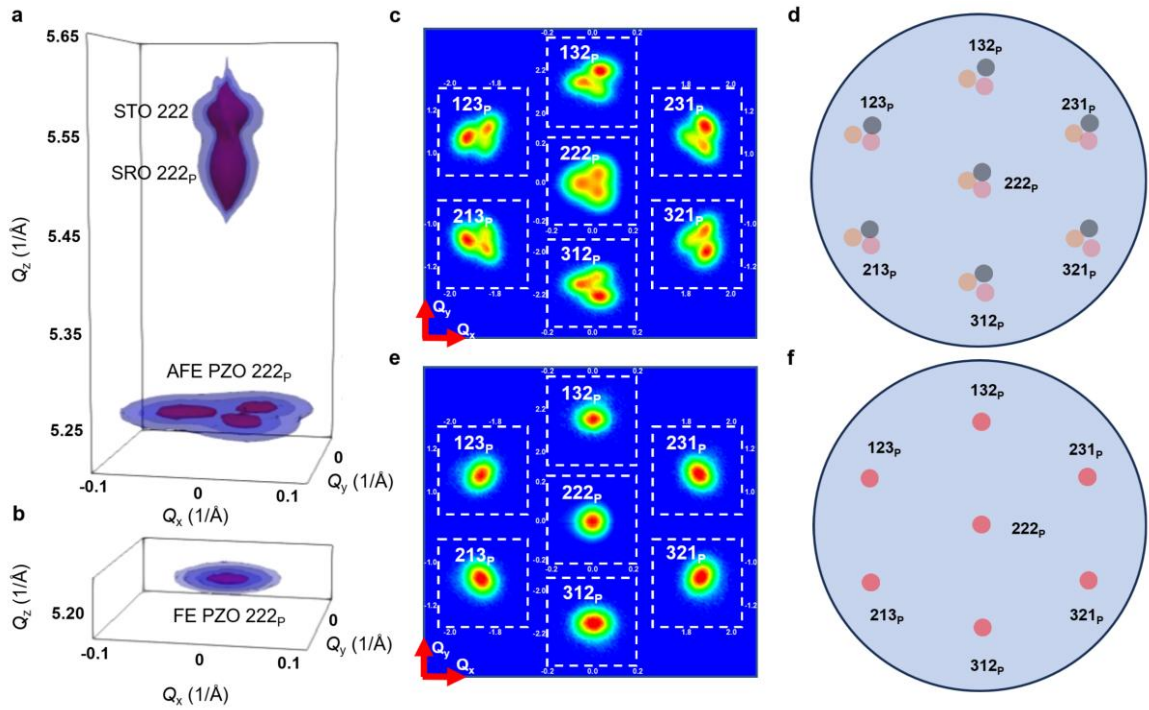

**Fig. 5 In-situ structure analysis of ground state AFE and field induced FE PbZrO$_3$. a,** Synchrotron-based 3D RSM around SrTiO$_3$ (222) for AFE PbZrO$_3$. $Q_x$//[11$\bar{2}$], $Q_y$//[1$\bar{1}$0], and $Q_z$//[111]. **b,** The 3D RSM of FE PZO 222 main spot. **c,** Wide range in-plane RSM of AFE PbZrO$_3$ at $Q_z$ = 5.24 Å$^{-1}$. **d,** Schematical view of diffraction pattern from different AFE domains. **e,** Wide range in-plane RSM of FE PbZrO$_3$ at $Q_z$ = 5.20 Å$^{-1}$. **f,** Schematical view of diffraction pattern from FE single domain.